# Origins of transverse voltages generated by applied thermal gradients and applied electric fields in ferrimagnetic-insulator/heavy-metal bilayers

Arnab Bose[1,2,*], Rakshit Jain[2,*], Jackson J. Bauer[3], Robert A. Buhrman[1], Caroline A. Ross[3], Daniel C. Ralph[2,4]

1. School of Applied and Engineering Physics, Cornell University, Ithaca, NY 14853, USA
2. Dept. of Physics, Cornell University, Ithaca, NY 14853, USA
3. Department of Materials Science and Engineering, Massachusetts Institute of Technology, MA 02139, USA
4. Kavli Institute at Cornell for Nanoscale Science, Ithaca, New York 14853, USA

ABSTRACT

We compare thermal-gradient-driven transverse voltages in ferrimagnetic-insulator/heavy-metal bilayers ($Tm_3Fe_5O_{12}$/W and $Tm_3Fe_5O_{12}$/Pt) to corresponding electrically-driven transverse resistances at and above room temperature. We find for $Tm_3Fe_5O_{12}$/W that the thermal and electrical effects can be explained by a common spin-current detection mechanism, the physics underlying spin Hall magnetoresistance (SMR). However, for $Tm_3Fe_5O_{12}$/Pt the ratio of the electrically-driven transverse voltages (planar Hall signal/anomalous Hall signal) is much larger than the ratio of corresponding thermal-gradient signals, a result which is very different from expectations for a SMR-based mechanism alone. We ascribe this difference to a proximity-induced magnetic layer at the $Tm_3Fe_5O_{12}$/Pt interface.

* Equal contributions

INTRODUCTION

In solid state materials that have a magnetic moment or are subject to an external magnetic field, an applied electric field or thermal gradient can generate transverse charge currents and voltages, resulting in Hall effects, anomalous Hall effects [1], or Nernst effects [2]. In systems with strong spin-orbit coupling, an electric field or thermal gradient can also drive analogous transverse-flowing spin currents, yielding spin Hall effects (SHE) [3] or spin Nernst effects (SNE) [4]. When heterostructures are made containing both layers with strong spin-orbit coupling and magnetic layers, spin currents can themselves generate transverse voltages and novel forms of magnetoresistance. The precise mechanisms by which these spin-current-driven electrical signals arise has been a matter of controversy, with arguments made for combinations of spin Hall magnetoresistance (SMR) [5,6] magnetic proximity effects (MPE) [7,8], and magnetic scattering [9]. Here we investigate these issues using a simple model system – combining a thin film of heavy metal with an insulating ferrimagnet ($Tm_3Fe_5O_{12}$ = TmIG), so that the transport characteristics are not affected by charge flow within the magnetic layer. We compare thermally-driven transverse voltages to their electrically-driven counterparts. We find that different mechanisms are active between TmIG/W and TmIG/Pt. In TmIG/W, both electrically-driven and thermally-driven transverse voltages can be understood from a single spin-current detection mechanism, the physics that gives rise to SMR. In TmIG/Pt, this is not the case, and we conclude that a MPE in the Pt as well as SMR affect the results.

BACKGROUND: SPIN HALL AND ANISOTROPIC MAGNETORESISTANCE

When an electric field ($E$) is applied to generate a charge current density $J_C$ within the plane of a ferrimagnetic insulator (FMI)/ heavy metal (HM) bilayer, a spin current density $J_S = \theta_{SH}(J_C \times \sigma)$ is created due to the spin Hall effect (SHE) in the HM (Fig. 1(a)), with $\theta_{SH}$ the spin Hall ratio of the HM and $\sigma$ the orientation of the current-generated spins being in-plane and perpendicular to $J_C$. A fraction of this spin current can be reflected at the interface depending upon the angle between $\sigma$ and the magnetization, $M$. The reflected spin current is then transduced back into an electric voltage within the HM by the inverse SHE. The result is spin Hall magnetoresistance (SMR) [5,6,10], a contribution to the sample resistance that depends on the magnetization orientation. For the definition of coordinate axes shown in Fig. 1(b), SMR produces changes in the longitudinal ($\rho_{xx}$) and transverse ($\rho_{xy}$) resistivity of the form:

$$\rho_{xx} = \rho_0 + \Delta\rho_{xy}^{\phi,SMR} m_y^2 \tag{1}$$

$$\rho_{xy} = \Delta\rho_{xy}^{z,SMR} m_z + \Delta\rho_{xy}^{\phi,SMR} m_x m_y \tag{2}$$

where $m_x = \sin\theta\cos\phi$, $m_y = \sin\theta\cos\phi$, and $m_z = \cos\theta$ represent the orientation of the magnetization saturated parallel to the applied magnetic field and $\Delta\rho_{xy}^{\phi,SMR}$ and $\Delta\rho_{xy}^{z,SMR}$ are the SMR coefficients. The values of $\Delta\rho_{xy}^{\phi,SMR}$ and $\Delta\rho_{xy}^{z,SMR}$ are predicted to depend on the real and imaginary parts of the effective spin mixing conductance of the interface ($G_{\text{eff}}^{\uparrow\downarrow}$) [6,10]:

$$\frac{\Delta\rho_{xy}^{\phi,SMR}}{\rho_0} = \theta_{SH}^2 \frac{\lambda}{t_{HM}} 2\lambda\rho_0 \text{Re}(G_{\text{eff}}^{\uparrow\downarrow}) \tag{3}$$

$$\frac{\Delta\rho_{xy}^{z,SMR}}{\rho_0} = -\theta_{SH}^2 \frac{\lambda}{t_{HM}} 2\lambda\rho_0 \text{Im}(G_{\text{eff}}^{\uparrow\downarrow}) \tag{4}$$

where for the heavy metal $\lambda$ is the spin diffusion length, and $t_{HM}$ is the thickness, and $G_{\text{eff}}^{\uparrow\downarrow} = G^{\uparrow\downarrow}\tanh^2\frac{t_{HM}}{2\lambda}/(1 + 2\lambda\rho_0 G^{\uparrow\downarrow}\coth\frac{t_{HM}}{\lambda})$ where $G^{\uparrow\downarrow}$ is the bare interfacial spin mixing conductance.

The existence of a spin Nernst effect (SNE) in ferromagnet/HM bilayers, a thermal analogue to the SHE, has been reported in Ref. [11–16]. In this case, an in-plane thermal gradient ($\nabla T_{ip}$) in a sample with no net charge current flow (i.e., open-circuit condition) generates a spin current density

$$J_S = -\theta_{SN}\frac{S_{HM}}{\rho_{HM}}(\nabla T_{ip} \times \sigma), \tag{5}$$

where $\theta_{SN}$ is the spin Nernst angle and $S_{HM}$ is the Seebeck coefficient of the HM. One can alternatively define a different quantity (we will call this $\theta_{SN}^0$) to characterize the spin current generated by a thermal gradient in a sample with no longitudinal electric field, so that in the presence of an electric field the total transverse spin current has the form

$$J_S = \theta_{SH}\frac{1}{\rho_{HM}}(E \times \sigma) - \theta_{SN}^0\frac{S_{HM}}{\rho_{HM}}(\nabla T_{ip} \times \sigma). \tag{6}$$

In the open-circuit condition corresponding to our measurements, there is a longitudinal electric field due to the Seebeck effect, $E = S_{HM}\nabla T_{ip}$, so that

$$\theta_{SN} = -\theta_{SH} + \theta_{SN}^0. \tag{7}$$

A thermally-generated spin current reflected at the interface can again be transduced into a voltage by the inverse-SHE in the heavy metal, resulting in thermally-induced voltages in both the longitudinal and transverse directions: [11–13]:

$$\frac{V_{xx}^{T-SMR}}{L_x} = -\left(S_{HM} + \Delta S_0^{SMR} + \Delta S_{xy}^{\phi,SMR}m_y^2\right)\nabla T_{ip} \tag{8}$$

$$\frac{V_{xy}^{T-SMR}}{L_y} = -\left(\Delta S_{xy}^{z,SMR}m_z + \Delta S_{xy}^{\phi,SMR}m_x m_y\right)\nabla T_{ip} \tag{9}$$

where $L_x$ and $L_y$ are the length and width of the device (Fig. 1(c)). Within the framework of the SMR spin-current detection mechanism, one would expect [10]

$$\frac{\Delta S_{xy}^{\phi,SMR}}{S_{HM}} = \theta_{SH}\theta_{SN}\frac{\lambda}{t_{HM}}2\lambda\rho_0\text{Re}(G_{\text{eff}}^{\uparrow\downarrow}) \tag{10}$$

$$\frac{\Delta S_{xy}^{z,SMR}}{S_{HM}} = -\theta_{SH}\theta_{SN}\frac{\lambda}{t_{HM}}2\lambda\rho_0\text{Im}(G_{\text{eff}}^{\uparrow\downarrow}). \tag{11}$$

If SMR is the only mechanism contributing to both the electrically-driven and thermally-driven voltage signals, then one should have

$$\frac{\Delta\rho_{xy}^{\phi,SMR}}{\Delta\rho_{xy}^{z,SMR}} = \frac{\Delta S_{xy}^{\phi,SMR}}{\Delta S_{xy}^{z,SMR}} \tag{12}$$

when all coefficients are measured at the same temperature. In this case, it should also be possible to determine $\theta_{SN}/\theta_{SH}$ by taking the ratio of either Eqs. (10) & (3) or Eqs. (11) & (4):

$$\theta_{SN}/\theta_{SH} = \frac{\Delta S_{xy}^{\phi,SMR}}{S_{HM}}\frac{\rho_0}{\Delta\rho_{xy}^{\phi,SMR}} = \frac{\Delta S_{xy}^{z,SMR}}{S_{HM}}\frac{\rho_0}{\Delta\rho_{xy}^{z,SMR}}. \tag{13}$$

In the case of an electrically-conducting magnetic film, such as Pt magnetized by the MPE, magnetization-dependent deflection of electrons can also produce both electrically and thermally-driven transverse voltage signals. In this case the longitudinal and transverse resistivities take the form [1]

$$\rho_{xx} = \rho_0 + \Delta\rho_{xy}^{\phi,AMR} m_x^2 \tag{14}$$

$$\rho_{xy} = \Delta\rho_{xy}^{z,AHE} m_z - \Delta\rho_{xy}^{\phi,AMR} m_x m_y \tag{15}$$

where $\Delta\rho_{xy}^{\phi,AMR}$ is the coefficient of anisotropic magnetoresistance (AMR) and $\Delta\rho_{xy}^{z,AHE}$ is the coefficient of the anomalous Hall effect (AHE). The thermal analogue for the transverse voltage contains terms corresponding to the anomalous Nernst effect (ANE) and planar Nernst effect (PNE)

$$\frac{V_{xy}^{T-AMR}}{L_y} = -\left(\Delta S_{xy}^{z,ANE} m_z - \Delta S_{xy}^{\phi,PNE} m_x m_y\right)\nabla T_{ip}. \tag{16}$$

Within an electrically-conducting magnet in general the equality analogous to Eq. (12) does not hold. For example, we show data for 1-nm-thick CoFeB samples in the Supplementary Materials [17] for which $\Delta S_{xy}^{\phi,AMR}$ is very weak, below our detection limit, so $\Delta\rho_{xy}^{\phi,AMR}/\Delta\rho_{xy}^{z,AHE}$ is at least a factor of 20 greater than $\Delta S_{xy}^{\phi,PNE}/\Delta S_{xy}^{z,ANE}$. This difference can be understood based on Mott's relation [2] assuming the scaling relationships $\Delta\rho_{xy}^{z,AHE} = \gamma_{AHE}\rho_{xx}^n$ and $\Delta\rho_{xy}^{\phi,PHE} = \gamma_{PHE}\rho_{xx}^m$. The electrically and thermally-driven coefficients for an electrically-conducting magnet should then be related as (see Supplementary Material):

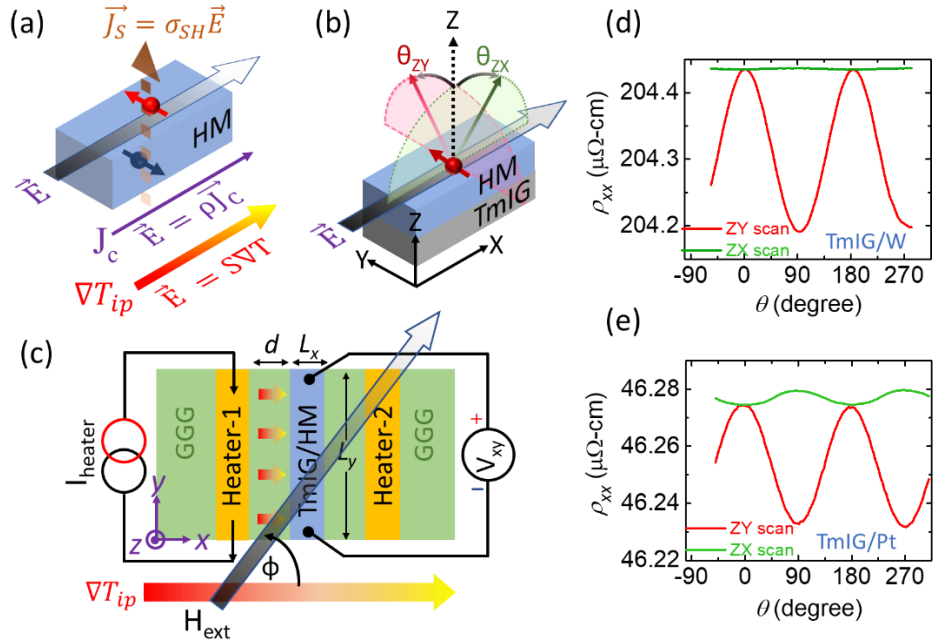

Fig. 1. (a) Origin of spin current generation by the spin Nernst effect. (b) Orientation of magnetic field rotation for *z-y* scans and *z-x* scans. (c) Schematic diagram of the experimental set-up for measurement of transverse voltages generated by a thermal gradient. (d,e) Longitudinal resistivities ($\rho_{xx}$) of the heavy-metal layer for (d) TmIG/W and (e) TmIG/Pt.

$$\frac{\Delta\rho_{xy}^{\phi,PHE}/\Delta\rho_{xy}^{z,AHE}}{\Delta S_{xy}^{\phi,PNE}/\Delta S_{xy}^{z,ANE}} = \frac{T\frac{\pi^2 k_B^2}{3e}\frac{\gamma'_{AHE}}{\gamma_{AHE}}-(n-1)S_{HM}}{T\frac{\pi^2 k_B^2}{3e}\frac{\gamma'_{PHE}}{\gamma_{PHE}}-(m-1)S_{HM}} \quad (17)$$

where $T$ is temperature, $k_B$ is the Boltzmann constant, $e$ is electronic charge, and $\gamma'_{PHE}$ and $\gamma'_{AHE}$ are the energy derivatives of $\gamma_{PHE}$ and $\gamma_{AHE}$. Since the mechanisms giving rise to the AHE and PHE are unrelated, in general $\gamma'_{AHE}/\gamma_{AHE} \neq \gamma'_{PHE}/\gamma_{PHE}$ and $n \neq m$. It follows that the ratio in Eq. (17) has no reason to be equal to 1 for an electrically-conducting magnetic layer.

EXPERIMENTAL METHODS

Experiments were performed on bilayers of TmIG with both W and Pt. The TmIG films (6 nm) were grown on single-crystal (111) gadolinium gallium garnet (GGG) substrates via pulsed laser deposition (PLD) in 150 mTorr O$_2$ at a substrate temperature of 650 °C from a bulk TmIG target [18]. The TmIG has perpendicular magnetic anisotropy (PMA) with an anisotropy field of 1.3 kOe, based on the applied field required to reorient the magnetization in-plane (see Supplemental Material). The PMA is a result of the in-plane tensile strain of the TmIG combined with its negative magnetostriction $\lambda_{111}$, which leads to a magnetoelastic anisotropy that overcomes the shape anisotropy.

After TmIG growth, the samples were transferred through air to a separate vacuum system where W (4 nm) or Pt (4 nm) was deposited by sputtering without further surface treatment. The average resistivities of the W and Pt films are 204 μΩ-cm and 46 μΩ-cm. The high resistivity in the W films indicates that they are primarily β-phase [19]. Device structures were patterned by optical lithography and then Ar-ion milling fully through the TmIG to the GGG substrate so that the only TmIG remaining on the sample chip is within the TmIG/HM wire. Electrical contacts were made by sputtering and lift-off of Ti (5 nm) and Pt (75 nm). More details about film growth, characterization, and sample fabrication are provided in the Supplemental Material.

We measure the longitudinal resistivity ($\rho_{xx}$) for both TmIG/W (Fig. 1(d)) and TmIG/Pt (Fig. 1(e)) while rotating the magnetic field (with magnitude 20 kOe, much higher than the anisotropy field) in the $z$-$x$ and $z$-$y$ planes (field orientations shown in Fig. 1(b)). Already we observe an important difference between the W and Pt samples. The TmIG/W devices show a large oscillation in $\rho_{xx}$ for magnetic field rotation in the $y$-$z$ plane with an angular dependence that fits well $\propto m_y^2$, with negligible variation of $\rho_{xx}$ for field rotation in the $x$-$z$ plane. This is consistent with a signal entirely due to SMR (Eq. (1)) with negligible AMR (Eq. (14)). In contrast, the TmIG/Pt sample has significant variation in $\rho_{xx}$ for field rotations in both the $x$-$z$ and $y$-$z$ planes, suggesting contributions from both SMR and AMR. Since TmIG is insulating, the presence of significant AMR in TmIG/Pt suggests the influence of a magnetic proximity effect in the Pt layer, as has been reported previously [20,21].

To measure the transverse voltages produced by a thermal gradient, we made TmIG/HM wires 650 μm long ($L_y$) and 20 μm wide ($L_x$) placed between two heater lines made of 15 nm thick and 200 μm wide Pt as shown in Fig. 1(c). The separation $d$ between the TmIG/HM wires and each heater line was varied in different devices between 15 μm and 200 μm. The advantage of this device geometry is that a $\nabla T_{ip}$ of either sign can be applied along the $x$-axis, and the large value of $L_y$ increases the magnitude of the measured thermally-induced voltage. We calibrated the temperature in the sample wires as a function of $d$ by measuring the resistances of both the heater wire and the sample wires when current is flowing in the heater, and comparing to independent measurements of resistance as a function of temperature using external heating of the entire sample chip. Based on these measurements we can map $\nabla T_{ip}$ as a function of heater spacing (Fig. 2(f)) (see Supplementary Information). We explored heater temperatures up to 400 K, which is significantly less than the 550 K Curie temperature of the TmIG.

MEASUREMENTS OF TRANSVERSE MAGNETORESISTANCE AND NERNST EFFECT FOR TmIG/W

Our measurements of the electric-field-driven Hall resistance for TmIG/W (Fig. 2(a,b)) as a function of magnetic field angle fit well to the dependence

$$R_{xy} = \Delta R_{xy}^z m_z + \Delta R_{xy}^\phi m_x m_y. \tag{18}$$

We determine $\Delta R_{xy}^z$ based on sweeps of field perpendicular to the sample plane (e.g., Fig. 2(a)), and $\Delta R_{xy}^\phi$ based on measurements as a function of rotating the field angle in the x-y plane (Fig. 2(b)). We have measured $\Delta R_{xy}^z$ and $\Delta R_{xy}^\phi$ from room temperature to 390 K as shown with the blue and red curves in Fig. 2(e). $\Delta R_{xy}^z$ is approximately independent of temperature while $\Delta R_{xy}^\phi$ decreases by almost a factor of two over this temperature range.

To measure the thermally-driven transverse voltage (Fig. 2(c,d)), we apply current through one of the Pt heaters adjacent to the TmIG/W wire, which generates thermal gradients both in the sample plane, $\nabla T_{ip}$, and out of the plane, $\nabla T_{op}$ at the position of the TmIG/W wire (Fig. 1 (a)). We then measure the transverse voltage in the TmIG/W wires generated as a function of changing the direction of the applied magnetic field (Fig. 2(c-d)). For a fixed heater power, we find that the transverse voltage as a function of magnetic-field angle is well described by

$$V_{xy}^T = \Delta V_{xy}^z m_z + \Delta V_{xy}^\phi m_x m_y + \Delta V_{SSE} m_x. \tag{19}$$

We measure $\Delta V_{xy}^z$ by sweeping the magnetic field perpendicular to the sample plane (Fig. 2(c)). $\Delta V_{xy}^\phi$ and $\Delta V_{SSE}$ are determined using measurements as a function of rotating the magnetic field in the sample plane with a fit to the function $\Delta V_{xy}^\phi \sin\phi\cos\phi + \Delta V_{SSE}\cos\phi$ (Fig. 2(d)). We ascribe the term proportional to $\cos\phi$ to the longitudinal spin Seebeck effect (SSE) [22] produced by $\nabla T_{op}$, that generates a transverse voltage by the inverse spin Hall effect (ISHE). As a function of heater power, $\Delta V_{xy}^z$ scales linearly (blue curve in Fig. 2(g)), while $\Delta V_{xy}^\phi$ scales slightly sub-linearly (red curve in Fig. 2(g)). This suggests that $\Delta S_{xy}^\phi \equiv \Delta V_{xy}^\phi/(L_x \nabla T_{ip})$ decreases with increasing temperature, similarly to $\Delta R_{xy}^\phi$. All three contributions, $\Delta V_{xy}^z$, $\Delta V_{xy}^\phi$, and $\Delta V_{SSE}$, decrease with increasing spacing between the heater and the TmIG/W wire (inset of Fig. 2(g), Fig. 2(h)) as expected on account of the decreasing thermal gradients, with the $\Delta V_{SSE}$ term (corresponding to $\nabla T_{op}$) decreasing fastest (Fig. 2(h)), such that the contribution from $\nabla T_{op}$ becomes negligibly small for spacings greater than 100 μm in the range of heater power we explored.

To test whether the mechanism behind SMR is sufficient to explain both the electrically-driven and thermally-driven transverse voltages, we consider the ratio

$$\eta = \left| \frac{\Delta R_{xy}^\phi / \Delta R_{xy}^z}{\Delta V_{xy}^\phi / \Delta V_{xy}^z} \right| = \left| \frac{\Delta \rho_{xy}^\phi / \Delta \rho_{xy}^z}{\Delta S_{xy}^\phi / \Delta S_{xy}^z} \right|. \tag{20}$$

As noted above (Eq. (12)), if SMR is the dominant spin-current detection mechanism contributing to both electrically-driven and thermally-driven signals, then $\eta$ is predicted to be equal to 1. In making this comparison, it is important to take into account that during the thermally-driven measurements the TmIG/W wires are heated substantially above room temperature (Fig. 2(g)), so we make the comparisons using values of $\Delta R_{xy}^{\phi}$ and $\Delta R_{xy}^{z}$ at the same elevated temperatures present for the thermally-driven measurements. The results are shown in Fig. 2(i). Over a range of heater spacing from 15 μm to 200 μm we find that the ratio $\eta$ is equal to one within 20%, which is on the scale of the experimental uncertainty in the measured values. We conclude that SMR is the dominant spin-current detection mechanism in determining both the electrically-driven and thermally-driven signals in TmIG/W.

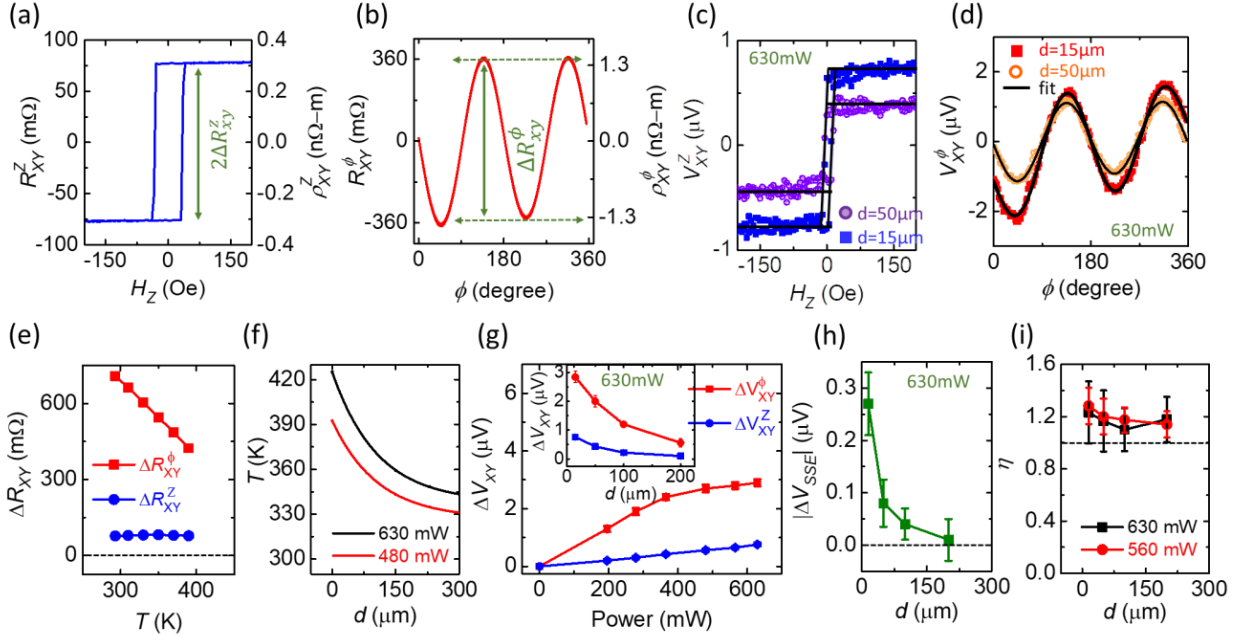

FIG. 2. Results for TmIG/W. Hall resistance of TmIG/W for (a) out-of-plane magnetic-field sweep and (b) in-plane field rotation for a field magnitude of 2.7 kOe. Thermally-induced transverse voltages, (c) $V_{xy}^{z}$ and and (d) $V_{xy}^{\phi}$, for a heater power of 630 mW and for two different heater spacings, $d$ = 15 μm (closed squares) and 50 μm (open circles). (e) Temperature dependence of $\Delta R_{xy}^{\phi}$ (red) and $\Delta R_{xy}^{z}$ (blue). (f) Temperature profile as function of $d$ for heater power, $P_{heater}$ = 630 mW (black) and 490 mW (red). (g) Heater power dependence of $\Delta V_{xy}^{\phi}$ (red) and $\Delta V_{xy}^{z}$ (blue) for $d$=15 μm. Inset of (g): Dependence on $d$ for $\Delta V_{xy}^{\phi}$ (red) and $\Delta V_{xy}^{z}$ (blue) for $P_{heater}$ =630 mW. (h) Spin-Seebeck component $\Delta V_{SSE}$ as function of $d$ for $P_{heater}$=630 mW. (i) Variation of $\eta$ as a function of $d$ for $P_{heater}$ = 630 mW (black) and 560 mW (red).

# MEASUREMENTS OF TRANSVERSE MAGNETORESISTANCE AND NERNST EFFECT FOR TmIG/Pt

Figure 3 shows results for the electrically-driven and thermally-driven transverse voltages for the TmIG/Pt samples analogous to those shown in Fig. 2 for TmIG/W. We observe both qualitative and quantitative differences. The signs of the electrically-driven coefficients $\Delta R_{xy}^{\phi}$ and $\Delta R_{xy}^{z}$ are the same for both heavy metals (compare Fig. 2(a,b) with Fig. 3(a,b)), but the signs of the thermally-driven signals $\Delta V_{xy}^{\phi}$ and $\Delta V_{xy}^{z}$ differ between TmIG/W and TmIG/Pt (compare Fig. 2(c,d) with Fig. 3(c,d)). There is also a striking difference in the scale of $\Delta R_{xy}^{\phi}$ and $\Delta R_{xy}^{z}$ for TmIG/Pt (compare Fig. 2(e) and Fig. 3(e)). The ratio $\Delta R_{xy}^{\phi}/\Delta R_{xy}^{z}$ varies from about 40 to 15 over the temperature range from room temperature to 400 K, whereas for TmIG/W this ratio varies only from about 4.6 to 2.5. Compared to the TmIG/W samples, $\Delta R_{xy}^{\phi}$ for TmIG/Pt is less dependent on the temperature (red curve in Fig. 3(e)) while $\Delta R_{xy}^{z}$ increases by factor of 2.7 (blue curve in Fig. 3(e)) from room temperature to 400 K rather than being approximately constant. As a function of heater power, both $\Delta V_{xy}^{\phi}$ and $\Delta V_{xy}^{z}$ for TmIG/Pt (red and blue curves in Fig. 3(g)) scale approximately linearly. Similar to TmIG/W, $\Delta V_{xy}^z$, $\Delta V_{xy}^z$ and $\Delta V_{SSE}$ all decrease as a function of heater spacing $d$ with $\Delta V_{SSE}$ decaying much faster (inset of Fig. 3(g), Fig. 3(h)).

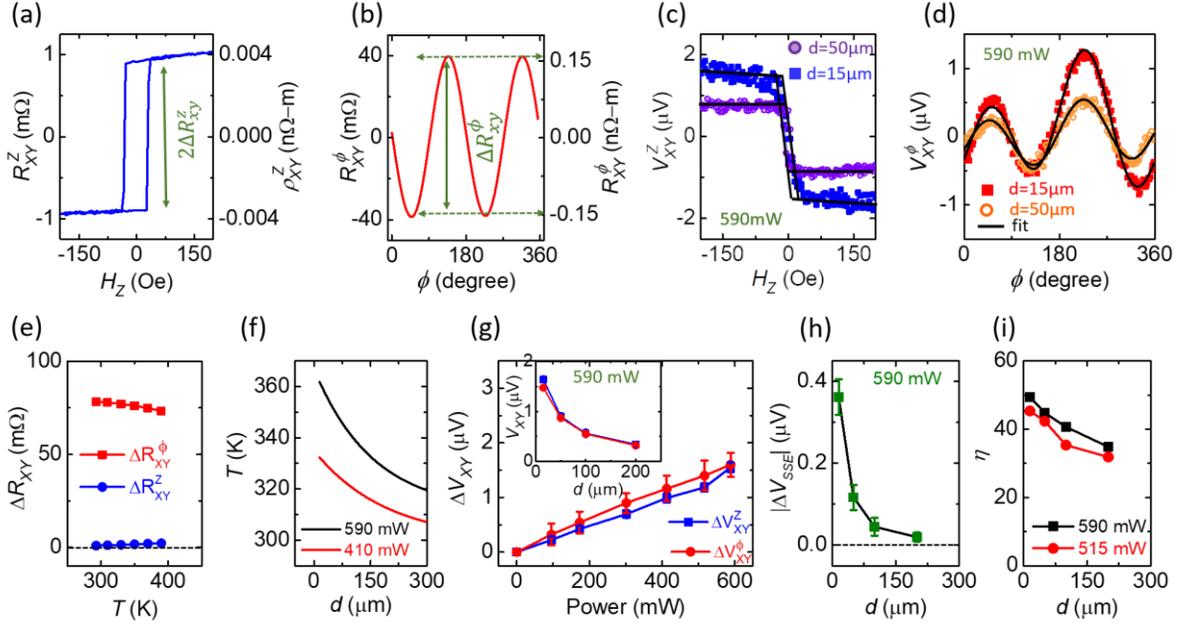

FIG. 3. Results for TmIG/Pt. Hall resistance of TmIG/Pt for (a) out-of-plane magnetic-field sweep and (b) in-plane field rotation for a field magnitude of 2.7 kOe. Thermally-induced transverse voltages, (c) $V_{xy}^{z}$ and and (d) $V_{xy}^{\phi}$, for a heater power of 590 mW and for two different heater spacings, $d$ = 15 μm (closed squares) and 50 μm (open circles). (e) Temperature dependence of $\Delta R_{xy}^{\phi}$ (red) and $\Delta R_{xy}^{Z}$ (blue). (f) Temperature profile as function of $d$ for heater power, $P_\text{heater}$ = 590 mW (black) and 410 mW (red). (g) Heater power dependence of $\Delta V_{xy}^{\phi}$ (red) and $\Delta V_{xy}^{z}$ (blue) for $d$=15 μm. Inset of (g): Dependence on $d$ for $\Delta V_{xy}^{\phi}$ (red) and $\Delta V_{xy}^{z}$ (blue) for $P_\text{heater}$ =590 mW. (h) Spin-Seebeck component $\Delta V_{SSE}$ as function of $d$ for $P_\text{heater}$=590 mW. (i) Variation of $\eta$ as a function of $d$ for $P_\text{heater}$ = 590 mW (black) and 515 mW (red).

Using the same technique we employed for TmIG/W, we can test for TmIG/Pt whether the spin-current detection mechanism associated with SMR is sufficient to explain both the electrically-driven and thermally-driven transverse voltages by calculating the ratio $\eta$ (Eq. (20)). For TmIG/Pt we find $\eta$ = 30-50 depending on the heater spacing $d$ (Fig. 3(i)), which is substantially different from our result $\eta \approx 1$ for TmIG/W (Fig. 2(i)). We therefore conclude for TmIG/Pt that the SMR spin-current detection mechanism cannot be dominant in determining both the electrically-driven and thermally-driven transverse voltages in this system. Based on previous measurements of a magnetic proximity effect in TmIG/Pt bilayers at room temperature [20], and our observation of an anisotropic magnetoresistance signal in our TmIG/Pt samples (Fig. 1(d)) consistent with a MPE in the Pt, we suggest that the difference between the TmIG/Pt and TmIG/W samples is the existence of an electrically-conducting magnetic layer in the Pt due to the proximity effect at the measurement temperature. The influence of a MPE in TmIG/Pt but not TmIG/W in our measurements is consistent with previous measurements that the onset temperature of the MPE upon cooling in TmIG(6 nm)/Pt is well above room temperature while in TmIG(6 nm)/W it is well below room temperature [20].

Previously, Avci et al. studied similar TmIG/Pt samples and also found values for $\eta \approx$ 8-10 [23], qualitatively similar to our result in that the value was much greater than 1. Instead of a MPE, they suggested that the mechanism for this result was a thermal spin-drag effect, in which the SSE of TmIG due to $\nabla T_{op}$ induces spin accumulation in the heavy metal, and $\nabla T_{ip}$ then induces an in-plane spin current that generates a transverse voltage due to the inverse-SHE. We can rule out this possibility for our samples because the dependence of our signals ($\Delta V_{xy}^z$) on heater power (Figs. 2(g), 3(g)) is to a good approximation linear, while $\nabla T_{op} \nabla T_{ip} \propto$ (heater power)$^2$. Furthermore, the out-of-plane thermal gradient $\nabla T_{op}$ should scale proportional to the spin Seebeck voltage $|\Delta V_{SSE}|$ as the heater spacing $d$ is varied (Fig. 3(h)). We find $\eta$ decreases much more slowly with increasing $d$ compared to $|\Delta V_{SSE}|$ (Figs. 3(h) & 3(i)), which again indicates that a thermal spin drag effect cannot explain our data.

DISCUSSION

Thus far we have focused on mechanisms by which the spin currents are detected, arguing that in TmIG/W both the electrically-driven and thermally-driven signals are consistent with a read-out mechanism dominated by spin Hall magnetoresistance (SMR), while this is not the case for TmIG/Pt. If we take as established that the thermally-generated transverse voltages for TmIG/W are due to SMR read-out, then our data also allow us to consider the mechanism by which the thermally-induced spin currents are generated in TmIG/W, by analyzing the value of the spin Nernst angle, $\theta_{SN}$ (Eqs. (5) & (7)).

If the transverse spin current generation in W were due entirely to the intrinsic SHE [3,24], theory predicts that one should be able to compute the transverse spin current as an appropriate integral over occupied electronic states of a transverse anomalous velocity

$$\vec{v}(\vec{k}, \vec{\sigma}) = e\vec{E} \times \vec{\Omega}(\vec{k}, \vec{\sigma}), \qquad (21)$$

where $\vec{k}$ is the electron wavevector, $\vec{\sigma}$ is a spin index, $e$ is the electron charge, $\vec{\Omega}(\vec{k}, \vec{\sigma})$ is the spin Berry curvature (see Ref. [25] for an elementary discussion) and $\vec{E}$ is the electric field. Since only the electric field and not a thermal gradient appears in Eq. (21), a thermal gradient should not directly generate a spin current by an intrinsic spin Hall effect. However, even in the absence of any net charge current a thermal gradient that gives rise to an electric field due to the Seebeck effect should generate the same spin current as if that value of electric field were applied externally. In other words, under the assumptions associated with Eq. (21), $\theta_{SN}^0$ in Eq. (6) should be zero and hence one should have $\theta_{SN} = -\theta_{SH}$ by Eq. (7).

Making a precise measurement of the absolute Seebeck coefficient of materials at room temperature is non-trivial because it is not possible to use a superconductor as a reference electrode. We have performed measurements on our TmIG(6 nm)/W(4 nm) films using lithographically-defined Au wires as reference electrodes and have corrected for the literature value of the absolute Seebeck coefficient of Au, ~ +1.5 µV/K [26] (see Supplemental Materials). The resulting estimate for the Seebeck coefficient for our 4 nm W films is $S_{HM}(W) = -4.5 \pm 0.5$ µV/K. With this value, Eq. (13) yields $\theta_{SN}/\theta_{SH} = -1.9 \pm 0.6$ based on $\frac{\Delta S_{xy}^{\phi,SMR}}{S_{HM}} \frac{\rho_0}{\Delta \rho_{xy}^{\phi,SMR}}$ and $\theta_{SN}/\theta_{SH} = -2.4 \pm 0.6$ based on $\frac{\Delta S_{xy}^{z,SMR}}{S_{HM}} \frac{\rho_0}{\Delta \rho_{xy}^{z,SMR}}$. These values are consistent with previous reports for bilayers consisting of W with a conducting magnet [12,13]. The fact that $\theta_{SN}/\theta_{SH}$ is of order $-1$ suggests to us that the intrinsic SHE largely sets the scale of the thermally-generated spin current in W. This is interesting in that even in the absence of any net charge current flow spin current is still driven by an electric field. However, we suggest that this is not the whole story, since $\theta_{SN}/\theta_{SH}$ differs from -1 by more than our estimated experimental uncertainty. Deviations might result from extrinsic contributions to the SHE or strong energy dependence of the spin Hall ratio.

SUMMARY

In conclusion, we find that the mechanisms that lead to the generation of transverse voltages are different between TmIG/W and TmIG/Pt samples. In TmIG/W, both the electrically-generated and thermally-generated transverse voltages are consistent with the spin-current detection mechanism associated with spin Hall magnetoresistance (SMR). For TmIG/Pt, in contrast, the ratio of the thermally-generated anomalous Nernst signal $\Delta S_{xy}^z$ to the corresponding anomalous Hall signal $\Delta \rho_{xy}^z$ is much larger than would be expected for a purely SMR-based spin-current detection mechanism. We suggest the reason for this difference is that proximity-induced magnetism exists near the TmIG/Pt interface in an electrically-conducting interface layer at room temperature, allowing for an anomalous Hall signal in addition to signals due to SMR.

ACKNOWLEDGEMENTS

A.B. originated the idea for the experiment and trained R.J. in experimental techniques. A.B. and R.J. then made equal contributions to the sample fabrication, measurements, and data analysis, supervised by D.C.R and R.A.B. J.J.B. grew the TmIG films, supervised by C.A.R. A.B was supported by the Cornell Center for Materials Research, funded by the National Science Foundation (NSF) MRSEC program, DMR-1719875. R.J. was supported by the US Department of Energy, DE-SC0017671. The samples were fabricated using the shared facilities of the Cornell NanoScale Facility, a member of the National Nanotechnology Coordinated Infrastructure (supported by the NSF, NNCI-2025233) and the facilities of Cornell Center for Materials Research. The research at MIT was supported in part by NSF Grant DMR-1808190 and SMART, an nCORE Center of the Semiconductor Research Corporation. Work at MIT made use of shared experimental facilities supported in part by the NSF MRSEC Program, DMR-1419807.

# Origins of transverse voltages generated by applied thermal gradients and applied electric fields in ferrimagnetic-insulator/heavy-metal bilayers

Arnab Bose, Rakshit Jain, Jackson J. Bauer, Robert A. Buhrman, Caroline A. Ross, Daniel C. Ralph

SUPPLEMENTARY INFORMATION

- A. Materials preparation
- B. Sample fabrication and measurement procedures
- C. Temperature calibration
- D. Measurement of the Seebeck coefficient
- E. Mott relation for the anomalous Hall effect and planar Hall effect
- F. Hf/CoFeB data
- G. Measurements for TmIG/W with reversed thermal gradient

## A. Materials preparation

The TmIG films were grown on (111) gadolinium gallium garnet (GGG) substrates via pulsed laser deposition (PLD) from a bulk TmIG target. A 248 nm KrF excimer laser was used to ablate the target at a fluence of 2.0 J/cm$^2$ and a 10 Hz repetition rate. The films were grown in 150 mTorr $O_2$ at a substrate temperature of 650 °C. Details of the sample growth and characterization can be found in [1]. The resulting films possessed perpendicular magnetic anisotropy (PMA) as confirmed in Magneto-Optic Kerr effect (MOKE) studies (Fig. S1(a)) and vibrating sample magnetometry (Fig. S1(b)) upon sweeping the magnetic field out of plane. This is consistent with the anomalous Hall measurements presented in the main text (Figs. 2(a),3(a)). To know the anisotropy field (required field to pull the magnet from out of plane to in-plane), an in-plane field is swept at an angle 45 degree with respect to the current flow direction while planar Hall voltage is measured (Fig. 1(c)). We find that anisotropy field is less than 1 kOe (Fig. 1(c)) while in the experiments presented in Fig. 2 and Fig. 3 of the main paper we applied 2.5 kOe magnetic field.

The Pt and W thin films are grown by dc sputtering at 30 W power and 2 mTorr Ar pressure after transferring TmIG films through the air. We did not perform any surface treatment before the deposition of Pt and W. At the thicknesses of 4 nm, the Pt and W films had electrical resistivities of 46 μΩ-cm and 204 μΩ-cm respectively. The high resistivity of the W films indicates that they are β-phase [2].

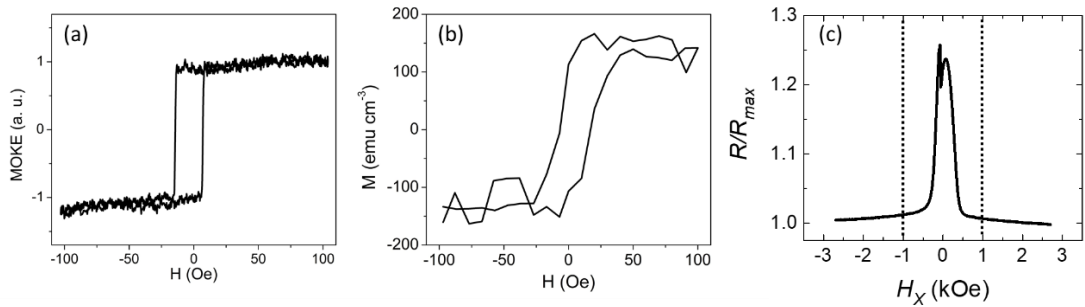

Fig. S1. (a) MOKE measurements of the magnetization of TmIG film as a function of magnetic field swept out-of-plane. (b) Vibrating sample magnetometry measurement as a function of out-of-plane magnetic field.

## B. Sample fabrication and measurement procedures

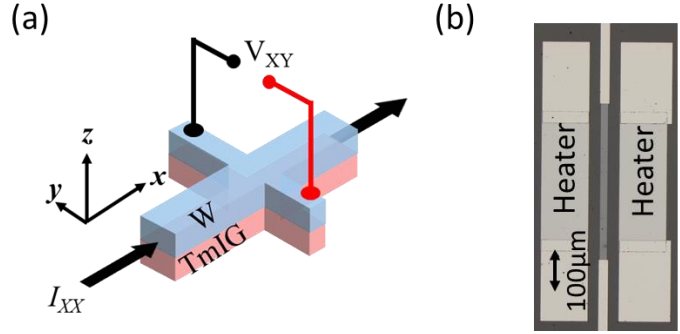

FIG. S2. (a) Schematic of the Hall bar for the electrically-generated anomalous Hall and planar Hall measurements. (b) Optical image of the heater lines for the thermally-generated transverse voltage measurements. The TmIG/W or TmIG/Pt wires lie in between the heaters at different spacings, $d$.

The Hall bars used for the longitudinal electrical measurements of the TmIG/W and (Fig. 1(d) in main text) and TmIG/Pt (Fig. 1(e) in the main text) had dimensions (130×5) µm² and (65×5) µm² respectively with 3-µm-wide side contacts. The Hall bars used for the transverse electrical measurements shown in Figs. 2 and 3 in the main text had dimensions (20×6) µm² with 4-µm-wide side contacts. To measure the electrically-generated transverse voltages we applied longitudinal charge current ($J_C$) on the order of $10^8$ A/m². We rotated an external magnetic field ($H_{ext}$) in the plane of the sample (*XY* plane) to measure $R_{xy}^{\phi}$, and swept $H_{ext}$ out-of-plane (along the *Z*-axis) to measure $R_{xy}^{z}$.

Figure S2(b) shows the device geometry for measurements of the thermally-generated transverse voltages. A TmIG/HM wire with dimensions 650×20 µm² is positioned between two large heaters. The heaters are made from Ti(5 nm)/Pt(15 nm) and each has dimensions 600×200 µm². The spacing between the TmIG/HM wire and the heaters had the values 15 µm, 50 µm, 100 µm, or 200 µm in different devices. The samples were made by first patterning the TmIG/HM wires by optical lithography and Ar ion etching through the TmIG to the substrate. The Pt heaters were then fabricated by optical lithography, sputtering, and lift-off. Finally Ta(5)/Pt(100) contact pads were deposited by the same technique. Electrical contact was made to the pads using Al wire bonds.

## C. Temperature calibration

For the TmIG/Pt sample, we applied fixed values of current to a heater and measured the resistance of both the heater and the TmIG/Pt wires at different distances $d$ from the heater ($d$ = 15, 50, 100, and 200 nm). These values were compared to measurements of resistance versus temperature upon heating the entire sample chip with an external heater, yielding values for temperature versus $d$ for fixed values of heater power. From these curves, the variation of $\nabla T_{ip}$ vs. $d$ was obtained by differentiating a smooth interpolation function.

For the TmIG/W sample, the TmIG/W wires themselves were not useful as local thermometers because of the weak dependence of the W resistance on temperature. Therefore, in this case we applied fixed values of current in a heater and measured the resistance in both the activated heater and other Ti/Pt heater lines on the other side of each of the TmIG/W wires, with heater-to-heater spacings of 60, 120, 220, and 420 nm. These values were again compared to measurements of resistance versus temperature upon heating the entire sample chip, and $\nabla T_{ip}$ vs. $d$ was obtained by differentiating a smooth interpolation function (Fig. S3). We assumed that the average temperature of each unheated wire corresponds to the temperature at the midpoint of the wire width. Figure S4 shows the measured curves of temperature versus distance and the resulting curves of $\nabla T_{ip}$ vs. $d$ for both the TmIG/W and TmIG/Pt wafers. In both cases the calculated in-plane thermal gradients are to good accuracy linear in the heater power (Fig. S4c,f).

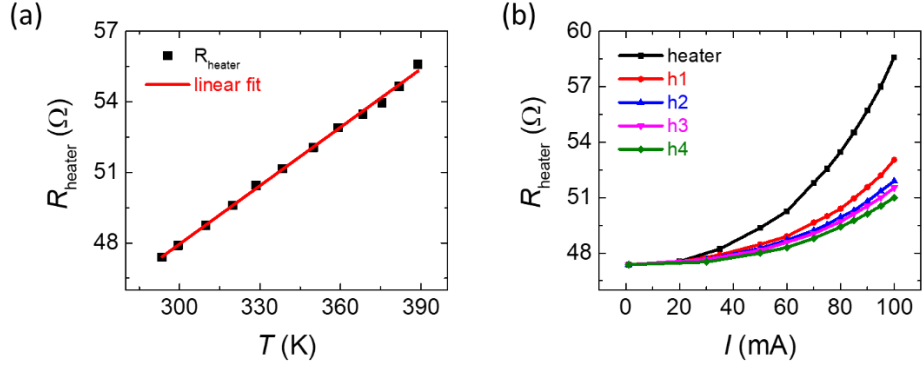

FIG. S3. (a) Resistance variation of the heater as a function of $T$ for the TmIG/W wafer. (c) Resistance variation of the activated heater (black curve) and the heater lines used as detectors on the TmIG/W after (red, blue, magenta and green curves) as a function of applied current. $I$=95 mA current corresponds to the maximum heater power which is 630 mW for TmIG/W devices.

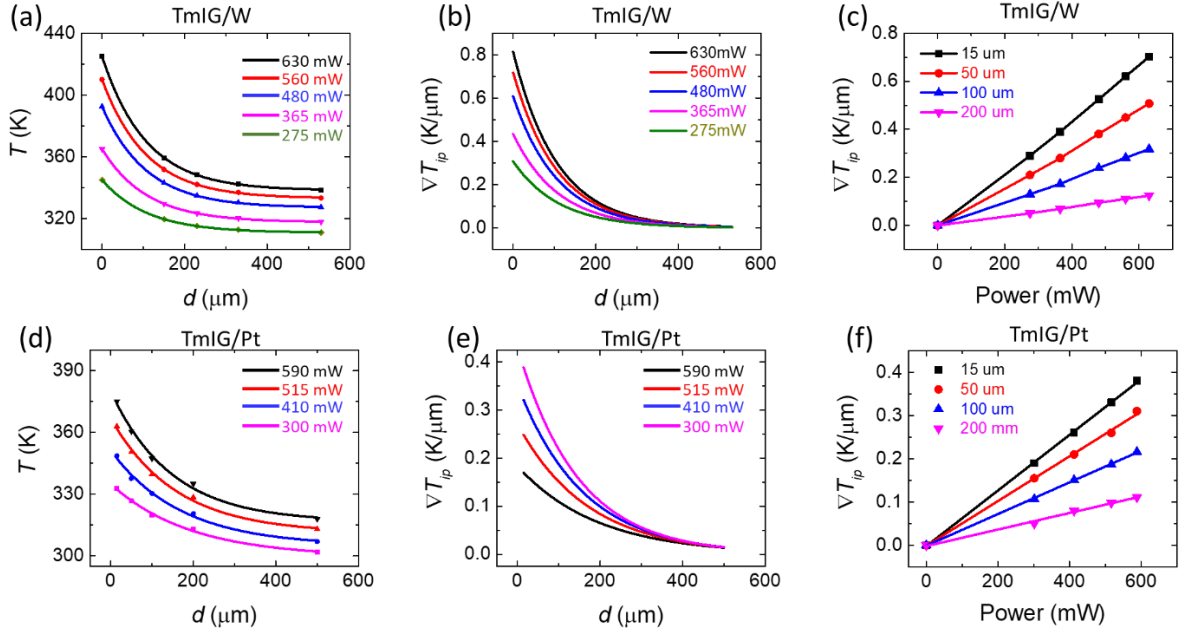

FIG. S4. Temperature profile for (a-c) TmIG/W and (d-f) TmIG/Pt. Local temperatures as a function of $d$ are shown in (a) and (d). The thermal gradients $\nabla T_{ip}$ calculated from these measurements are shown in (b) and (e). (c,f) $\nabla T_{ip}$ is to good accuracy a linear function of applied heater power.

## D. Measurement of the Seebeck coefficient

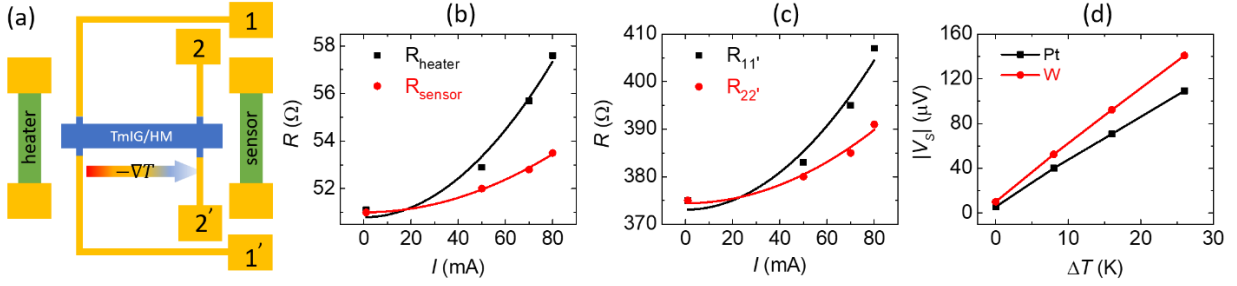

Fig. S5. (a) Schematic of the device used to measure the Seebeck coefficient. (b) Resistance of the heater (black) and the sensor (red) as a function of heater current. (c) Resistance of the transverse leads near the heater (black) and near the sensor (red). (d) Measured Seebeck voltage as a function of estimated temperature difference (ΔT) between two terminals (1(hot) and 2(cold)).

To make an estimate of the Seebeck coefficient in our thin films of Pt and W, we fabricate sample structures as shown in Fig. S5(a). A bar of TmIG(6 nm)/Pt(4 nm) or TmIG(6 nm)/W(4 nm) is patterned 2 mm long by 0.2 mm wide. Four contact wires made from Ta(5 nm)/Au(100 nm) are attached to each bar (with two near either end as shown in the figure), with contact pads 1, 1', 2, and 2' positioned close together on one side a sample (opposite to a heater) so that all of the contact pads will remain at a similar temperature throughout heating measurements. We choose Au for the contact wires because the absolute Seebeck coefficient of Au is expected to be small [3] and we calculate the Seebeck coefficients of our Pt and W films relative to this value. The Au layer in the contact wires is also sufficiently thick and wide that the resistance of the contact wires (< 10 Ω) is negligible compared to resistance of the TmIG/HM devices in the transverse direction (~ 2 kΩ for W and ~ 400 Ω for Pt). Therefore, temperature-dependent variations in resistance between probes 1 and 1', and between 2 and 2', are sensitive primarily to the temperature-dependent local resistance of the heavy-metal layer between the contact points. The sample structure is completed by heater and sensor wires made from Ta(3nm)/Pt(12 nm) with dimensions (0.5×0.2) mm$^2$ positioned 15 μm from the end of each TmIG/HM bar.

During measurements, we apply varying amounts of direct current through the heater to establish a thermal gradient along the TmIG/HM bar. While the current is applied, we measure the resistance of both the heater and the sensor (Fig. S5(b)). To determine the temperatures at the two ends of the TmIG/Pt bar during heating, we measure the resistances the transverse leads near the hot end of the TmIG/Pt bar ($R_{11'}$) and cold end ($R_{22'}$) (Fig. S5(c)) and compare to calibrations of the resistances as a function of temperature upon heating the entire sample chip with an external heater. From this we determine the temperature difference $\Delta T$. The thermoelectric voltage $V_S$ is measured as the voltage of contact pad 1 relative to contact pad 2 (Fig. S5(d). The Seebeck coefficient is then estimated as $S_{Pt} = -V_S / \Delta T + 1.5$ μV/K, after correcting for the contribution from the Au leads. The absolute Seebeck coefficient of our thin-film Pt is positive, so the magnitude of our final result for Pt is increased by this correction. For our TmIG(6 nm)/Pt(4 nm) layers we determine $S_{Pt} \approx +4.8$ μV/K.

The temperature dependence of 4-nm-thick β-phase W is sufficiently weak that we cannot make accurate determination of the local temperature of the TmIG/W bar by using measurements of $R_{11'}$ and $R_{22'}$. Instead, for the W sample we assume that the spatial dependence of the temperature profile within the GGG substrate is the same as for the TmIG/Pt sample. We measure the total temperature difference between the heater and the sensor on the TmIG/W sample chip and calculate the temperature difference between the two ends of the TmIG/W bar as the same fraction of the temperature difference between the heater and sensor as measured for the TmIG/Pt sample bar. The thermoelectric voltage we determine for TmIG(6 nm)/W(4 nm) is $S_W \approx -5.2$ μV/K. Both this value and our value for TmIG(6 nm)/Pt(4 nm) have signs opposite to the bulk Seebeck coefficients for W and Pt.

### E. Mott relation for the anomalous Hall effect and planar Hall effect

The transverse Seebeck coefficients can be expressed as [4]:

$$S_{xy} = \frac{1}{\sigma_{xx}}\left(\alpha_{xy} - \sigma_{xy}S_{xx}\right) \tag{1}$$

where $\sigma_{xx}$ is the longitudinal electrical conductivity ($\sigma_{xx} = \frac{1}{\rho_{xx}}$), $S_{xx}$ is longitudinal Seebeck coefficient equal by the Mott relation to

$$S_{xx} = \frac{\pi^2 k_B^2 T}{3e\sigma_{xx}}\left(\frac{\partial \sigma_{xx}}{\partial E}\right)_{E_F} \tag{2}$$

and $\alpha_{xy}$ is the Nernst conductivity:

$$\alpha_{xy} = \frac{\pi^2 k_B^2 T}{3e}\left(\frac{\partial \sigma_{xy}}{\partial E}\right)_{E_F} \tag{3}$$

Here $E$ is energy, $E_F$ is the Fermi energy, $T$ is temperature, $e$ is the electronic charge, and $k_B$ is the Boltzmann constant.

Given that the transverse resistivity and Seebeck coefficient depend on the orientation of the magnetization $\hat{m}$ as

$$\rho_{xy} = \Delta\rho_{xy}^{z,AHE} m_z - \Delta\rho_{xy}^{\phi,AMR} m_x m_y \tag{4}$$

$$S_{xy} = \Delta S_{xy}^{z,ANE} m_z - \Delta S_{xy}^{\phi,PNE} m_x m_y \tag{5}$$

and using the phenomenological scaling relationships

$$\left.\begin{array}{l}\Delta\rho_{xy}^{z,AHE} = \gamma_{AHE}\rho_{xx}^n \\ \Delta\rho_{xy}^{\phi,AMR} = \gamma_{PHE}\rho_{xx}^m\end{array}\right\} \tag{6}$$

with some algebra one can show in the regime that $\Delta\rho_{xy} \ll \rho_{xx}$ (so that $\sigma_{xy} \approx -\Delta\rho_{xy}/\rho_{xx}^2$) that

$$\Delta S_{xy}^{z,ANE} = \frac{\Delta\rho_{xy}^{z,AHE}}{\rho_{xx}}\left(T\frac{\pi^2 k_B^2}{3e}\frac{\gamma'_{AHE}}{\gamma_{AHE}} - (n-1)S_{xx}\right) \tag{7}$$

$$\Delta S_{xy}^{\phi,PNE} = \frac{\Delta\rho_{xy}^{\phi,AMR}}{\rho_{xx}}\left(T\frac{\pi^2 k_B^2}{3e}\frac{\gamma'_{PHE}}{\gamma_{PHE}} - (m-1)S_{xx}\right). \tag{8}$$

As a consequence,

$$\eta = \frac{\Delta\rho_{xy}^{\phi,PHE}/\Delta\rho_{xy}^{z,AHE}}{\Delta S_{xy}^{\phi,PNE}/\Delta S_{xy}^{z,ANE}} = \frac{T\frac{\pi^2 k_B^2 \gamma'_{AHE}}{3e\ \gamma_{AHE}} - (n-1)S_{HM}}{T\frac{\pi^2 k_B^2 \gamma'_{PHE}}{3e\ \gamma_{PHE}} - (m-1)S_{HM}}. \tag{9}$$

## F. Electrically-generated and thermally-generated transverse voltages for Hf(2 nm)/CoFeB(1 nm)

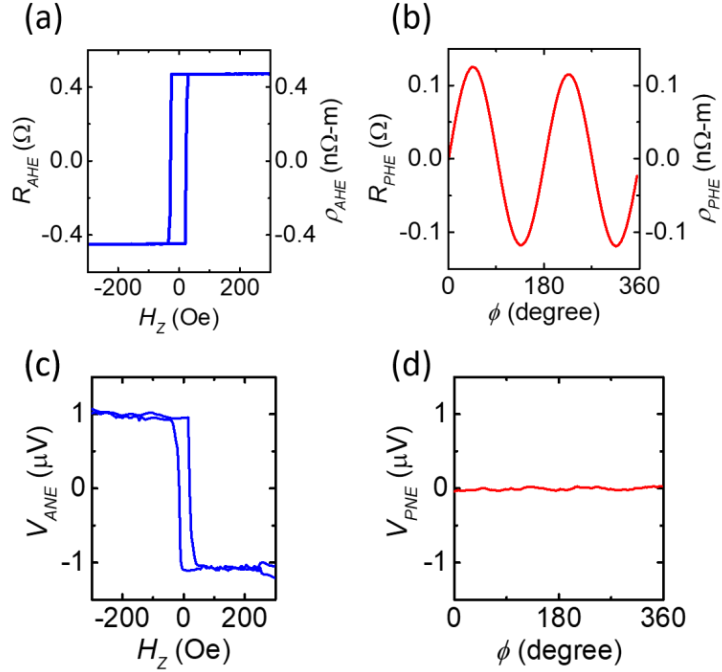

FIG. S6. Signals from a Hf(2 nm)/CoFeB(1 nm) sample with a total resistance approximately 800 Ω. (a) Anomalous Hall effect as a function of magnetic field swept out-of-plane. (b) Planar Hall effect as a function of angle for an in-plane magnetic field. (c) Anomalous Nernst effect as a function of magnetic field swept out-of-plane. (d) Planar Nernst effect as a function of angle for an in-plane magnetic field.

## G. Measurements for TmIG/W with reversed thermal gradient

We have performed measurements with the in-plane temperature gradient reversed by using heater 2 rather than heater 1 (see the diagram in Fig. S7(a)). When the thermal gradient is reversed we find (as expected) that the transverse voltage signals change sign for both the out-of-plane magnetic field scan (Fig. 7(b)) and the scan versus in-plane magnetic-field angle (Fig. 7(c)), while the measured magnitudes are nearly unchanged. This is additional evidence that the transverse voltage signals are due to the in-plane thermal gradient, and not the out-of-plane thermal gradient.

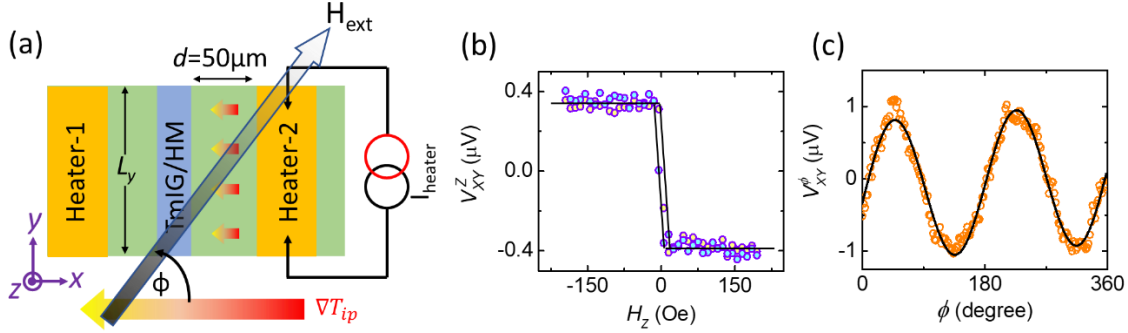

Fig. S7. (a) Schematic of the measurement in which the in-plane temperature gradient is reversed by applying current to heater 2 rather than heater 1. (b,c) Measurements of transverse voltage in this configuration for a TmIG/W sample with spacing $d$=50 μm and heater power 630 mW: (b) transverse voltage as a function of out-of-plane magnetic field and (c) as a function of in-plane field rotation for a field magnitude of 2.7 kOe.